\documentclass[preprint,superscriptaddress,nofootinbib]{revtex4}
\pdfoutput=1
\usepackage{bbm}
\usepackage{amsfonts}
\usepackage{mathrsfs}

\usepackage{amsmath,amssymb}
\usepackage{setspace}
\usepackage{epsfig}
\usepackage{graphicx}               % Standard graphics package
\usepackage{url}
\usepackage{hyperref}
\usepackage{float}
\usepackage{pstricks}
\usepackage{color}
\usepackage{slashed}
\usepackage{ulem}
\usepackage[caption=false]{subfig}

\newcommand{\nc}{\newcommand}

\nc{\beq}{\begin{equation}} \nc{\eeq}{\end{equation}}
\nc{\beqa}{\begin{eqnarray}} \nc{\eeqa}{\end{eqnarray}}
\nc{\bea}{\begin{eqnarray}} \nc{\eea}{\end{eqnarray}}
\nc{\barray}{\begin{eqnarray}} \nc{\earray}{\end{eqnarray}}
\nc{\barrayn}{\begin{eqnarray*}} \nc{\earrayn}{\end{eqnarray*}}
\nc{\ra}{\rightarrow}
\newcommand{\lsim}{\!\mathrel{\hbox{\rlap{\lower.55ex \hbox{$\sim$}} \kern-.34em
 \raise.4ex \hbox{$<$}}}}
\newcommand{\gsim}{\!\mathrel{\hbox{\rlap{\lower.55ex \hbox{$\sim$}} \kern-.34em
 \raise.4ex \hbox{$>$}}}}
\nc{\Tr}{{\rm Tr}} \nc{\slsh}{\slash\hspace*{-0.22cm}}

\def\be{\begin{equation}}
\def\ee{\end{equation}}
\def\bea{\begin{eqnarray}}
\def\eea{\end{eqnarray}}
\def\bit{\begin{itemize}}
\def\eit{\end{itemize}}

\nc{\infinity}{\infty} \nc{\mc}{\mathcal} \nc{\M}{\mathcal{M}}

\def\to{\rightarrow}

\newcommand{\lambdac}{\Lambda_{C}}

\begin{document}

\setlength{\baselineskip}{0.22in}

\begin{flushright}MCTP-16-16 \\
\end{flushright}
\vspace{0.2cm}

\title{Naturalness from a Composite Top?}
\author{Aaron Pierce and Yue Zhao}
\vspace{0.2cm}
\affiliation{Michigan Center for Theoretical Physics (MCTP) \\
Department of Physics, University of Michigan, Ann Arbor, MI
48109}

\date{\today}

\begin{abstract}
We consider a theory with composite top quarks but an elementary
Higgs boson.  The hierarchy problem can be solved by supplementing
TeV scale top compositeness with either supersymmetry or Higgs
compositeness appearing at the multi-TeV scale. The Higgs boson
couples to uncolored partons within the top quark.  We study how
this approach can give rise to a novel screening effect that suppresses
production of the colored top partners at the LHC. Strong
constraints arise from $Z$ to $\bar{b} b$, as well potentially from
flavor physics.  Independent of flavor considerations, current
constraints imply a compositeness scale $\gsim$ TeV; this implies
that the model is likely tuned at the percent level.  Four top quark
production at the LHC is a smoking-gun probe of this scenario. New
CP violation in D meson mixing is also possible.
\end{abstract}

\maketitle

\section{Introduction}
The large top quark Yukawa coupling $y_t \approx 1$ produces the
dominant corrections to the Standard Model (SM) Higgs boson
(mass)$^{2}$: $\Delta m_h^{2} \sim \frac{3y_t^2}{4\pi^2} \Lambda^2$.
Here, $\Lambda$ is a UV cut-off scale at which new physics appears
to cancel the quadratic divergence. If $\Lambda$ is much larger than
the electroweak (EW) scale, these large quantum corrections combine
with a nearly equal and opposite bare contribution to yield the much
smaller weak scale \cite{Susskind:1978ms}.  This apparent conspiracy
is known as the fine-tuning or gauge hierarchy problem.

In most solutions to the hierarchy problem, colored particles near
the weak scale provide the necessary cutoff $\Lambda$.  For
example, in supersymmetry (SUSY) the superpartner of top quark does the job. In this paper, we instead assume that the third
generation of quarks are composite particles that emerge after the
confinement of a strongly coupled gauge group, which happens the TeV scale. Contrary to most theories of strong
dynamics, we assume the Higgs boson is an elementary scalar and
couples to some of the color neutral partonic constituents of the top quark.  The
top Yukawa coupling is an induced coupling from the low-energy
effective theory point of view. This setup separates the Yukawa
coupling (responsible for the Higgs mass correction), from the
$SU(3)_c$ charge (potentially a source of stringent collider
constraints).

Composite top quarks composite do not stabilize the Higgs
(mass)$^{2}$ by themselves.  Above the compositeness scale, a
quadratic contribution to the elementary Higgs (mass)$^2$ remains,
arising from Yukawa interactions with the color neutral partons. An
additional mechanism -- such as SUSY at a slightly higher energy
scale -- is needed to cancel this contribution. This additional
mechanism likely introduces colored top partners after the strong
dynamics confines, but we will show hadronization under the new
strong dynamics suppresses the production of those heavy $SU(3)_c$
colored states.

\begin{figure}
\includegraphics[width=0.5\textwidth]{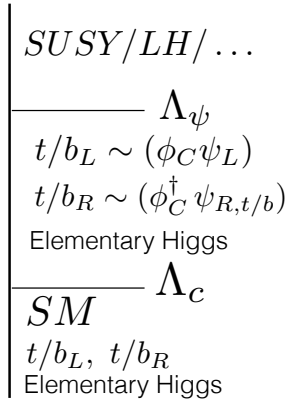}
\caption{Above $\lambdac$ the top/bottom quark is revealed to be composites.
Above $\Lambda_{\psi}$ additional physics enters (e.g. supersymmetry or a Little Higgs model) to eliminate remaining quadratic divergences to the Higgs boson mass.}
\label{fig:idea}
\end{figure}

In our model, the 1-loop quadratically divergent correction to Higgs
mass can be split into two pieces:
\begin{eqnarray}
 \label{eq:QuadHiggs}
\delta m_H^2\simeq \frac{3y_t^2}{4\pi^2} \lambdac^2+ \frac{N
y_\psi^2}{4\pi^2} \Lambda_{\psi}^2.
\end{eqnarray}
Here, $\lambdac$ is the confinement scale of our new strong gauge
interaction, and $\Lambda_{\psi}$ is the UV scale where additional
(presumably colorless) new physics should be introduced to cancel
the corrections from the colorless partons $\psi$ to the Higgs mass.  The basic idea is summarized in Fig.~\ref{fig:idea}.
We will find consistency with current bounds from precision
electroweak constraints and collider searches requires $\lambdac
\gsim$ TeV. The strongly coupled nature of the theory
introduces uncertainty in this estimate.

The idea of composite top quarks has a long history
\cite{Georgi:1994ha}, with recent discussions in
\cite{Pomarol:2008bh}, and emphasis on collider signals in
\cite{Lillie:2007hd,Kumar:2009vs,Fabbrichesi:2013bca,Englert:2014oea}.
There are also a host of theories where the Higgs boson is
composite, and the top is largely composite \cite{Kaplan:1991dc},
for reviews, see \cite{Contino:2006nn,Panico:2015jxa}. In our
set-up, however, we imagine that the Higgs boson is still an
elementary scalar. This impacts the pattern of low-energy deviations
from the Standard Model as well as the way in which the model should
be UV completed above the compositeness scale.

In the following section, we discuss the relationship between the
partonic Yukawa coupling and the induced effective Yukawa coupling
between the Higgs and the top quark. We will see the partonic Yukawa
coupling might even be ${\mathcal O}(0.1)$ while remaining
consistent with an ${\mathcal O(1)}$ top Yukawa coupling. We then
discuss an important constraint on the confining theory:
confinement must occur without chiral symmetry breaking. If this
were not the case, the top quark would get a too large mass. In Sec.
\ref{sec:UVcompletion}, we discuss UV models which may be present above the confinement scale $\Lambda_C$ to truncate the remaining quadratic divergences from the Yukawa couplings between the Higgs and color neutral partons. In Sec.
\ref{sec:Screening}, we note how the hadronization of the new strong
interaction can dramatically reduce the production of heavy
$SU(3)_c$ colored composite states with mass above $\Lambda_C$.
Although the existence of these states are unavoidable, this
screening effect removes or delays the appearance of
these particles at a hadron collider. Then, in
Sec.~\ref{sec:constraints}, we review phenomenological bounds on our
scenario, both from low-energy probes and from direct probes of
compositeness at the LHC. Some additional flavor constraints appear
in an Appendix. In Sec.~\ref{sec:collider}, we discuss the collider
signatures of this model as energies approach the compositeness
scale. In another appendix, we make some comments regarding anomaly
cancelation and how this module might be embedded in a UV theory.

\section{Induced top Yukawa}

The Yukawa coupling between the top quark and Higgs boson is induced by
the Yukawa coupling between colorless partons $\psi_{L/R}^\alpha$ within the top
and the Higgs boson, i.e. $y_{\psi}H\psi_L\psi_R$. The form factors translating the partonic
coupling $y_\psi$ to the bound state coupling $y_t$ are not known.
One might worry whether the partonic Yukawa coupling needs to
be very large to achieve a ${\mathcal O} (1)$ top Yukawa
coupling. If that were the case, it would not be attractive from a
naturalness point of view: the quadratic divergence from
the partons would become larger than that from top quark, see
Eq.~(\ref{eq:QuadHiggs}).  The relationship between the Yukawa
couplings can be estimated by na\"ive dimensional analysis (NDA)
\cite{NDA1, NDA2}, see also \cite{Jenkins:2013sda}. The effective
Lagrangian can be written as
\begin{eqnarray}
 \label{NDAYukawa}
\mathscr{L}_{NDA}\supset f^2\lambdac^2\bigg(\frac{y_\psi
H}{\lambdac}\bigg)\bigg(\frac{Q}{f\sqrt{\lambdac}}\bigg)\bigg(\frac{t_R}{f\sqrt{\lambdac}}\bigg)=y_\psi
H Q t_R.
\end{eqnarray}
Here, $\lambdac$ represents the scale of compositeness -- the confinement scale of the new strong dynamics.
%Here,
We take the dimensionful parameter associated with Higgs as
$\lambdac$ instead of $f$ (as is done in composite Higgs scenarios).
This is because the Higgs is an elementary particle that does not
participate in the strong dynamics. $\lambdac \equiv g_{\ast} f$,
where $g_{\ast}$ is a typical strong coupling.  In NDA, it is taken
to be $g_{\ast} \simeq 4 \pi$.  (We define $g_{\ast}$ as having no N-dependence; we will discuss the subtlety of N-dependence later.)  So, NDA predicts $y_t \simeq y_{\psi}$. Although NDA only provides guidance
on the magnitude of couplings in the low-energy effective theory, it at least indicates the Yukawa coupling between the Higgs and top quarks is not dramatically suppressed.

We can potentially gain further insight into the translation between
the parton-level and low-energy effective theory Yukawa couplings by studying a similar
scenario in the SM.  There, we can discuss how the coupling of the Higgs
boson to the proton relates to the underlying couplings between the
Higgs and the quarks.  These form factors are well-studied, e.g. in
the context of dark matter direct detection, and we quote the
results \cite{Cheng:1988im,Belanger:2013oya}:
\begin{eqnarray}
 \label{eqn:FF1}
B^{n,p}_{u,d}& \equiv &\frac{y_{u,d}}{y^{n,p}_{u,d}}\sim {\mathcal O}(0.1), \label{eqn:updown}\\
\label{eqn:FF2}
B^{n,p}_{s}&\equiv &\frac{y_{s}}{y^{n,p}_{s}}\sim {\mathcal O}(1).
\end{eqnarray}
We have defined $y^{n,p}_{q}$ in terms of the usual nucleon parameters as: $y^{n,p}_q \equiv m_{N} f^{T}_{q} /v$, where $v$ is the electroweak vacuum expectation value (VEV) and $m_{N}$ the nucleon mass.  We emphasize  these equations describe the relationship between the quark Yukawa couplings and the induced Higgs-nucleon-nucleon coupling.\footnote{In the SM, the Higgs-nucelon-nucleon coupling receives a dominant contribution from the Higgs-gluon-gluon coupling after integrating out heavy quarks.  In our set-up, we expect the top Yukawa to predominately arise from the Higgs-parton coupling.}  The results of Eqs.~(\ref{eqn:FF1}),(\ref{eqn:FF2}) are roughly consistent with NDA expectations.

It is not unreasonable that the form factors may in fact deviate from one. Scalar field interactions with
sea-quarks add; there is no cancellation between particle and
anti-particle. So, it may be that $y_{\psi}$ smaller than $y_t$, say $y_\psi\sim 0.1$. This would
effectively postpone the need for new states which cut off
divergences from the Higgs coupling to $\psi_{L/R}$.

\section{Confinement without chiral condensation}

The partons comprising the top quarks are assumed to be fermionic
and massless. Their masses are protected by chiral symmetry.
However, this is insufficient to ensure top quarks remain
massless after confinement. Indeed, after QCD confines, the only
light hadrons with $m << \Lambda_{QCD}$ are pions and kaons, which
are pseudo-Nambu-Goldstone bosons from the spontaneously broken
approximate chiral symmetry. There are no light fermions after QCD
confines. This is because chiral condensation occurs at a similar
scale to QCD confinement, $\langle \bar q q\rangle \sim 4 \pi
f_{\pi}^3$.  If our strong dynamics were to simultaneously confine
and break chiral symmetry, we would expect the top quark would get
mass at the confinement scale, just like the proton of QCD.  This would mean a too large top mass of ${\mathcal O}$(TeV).

But chiral symmetry breaking need not happen when the gauge group
confines. SUSY QCD provides an existence proof: for an $SU(N_c)$
gauge group with $N_F=N_C+1$, there are massless fermions at the
origin of the moduli space of the dual theory.  Those massless
fermions are bound states of the elementary particles in the theory
before duality \cite{Seiberg:1994bz}.  Confinement without chiral
condensation has also been studied in early attempts to explore the
SM as a relic a strongly coupled theory, see,
e.g.~\cite{Raby:1979my,Dimopoulos:1980hn}.

While we are agnostic as to the identity of the new strong gauge
group, for concreteness we refer to it as $SU(N)$.  We assume $SU(N)$ confinement occurs without chiral
condensation, and the top quark mass comes from the VEV of the Higgs alone.  The  massless composite fermions
are written as bound states of a scalar boson $\phi_c$ and a fermion
$\psi_{L/R}^\alpha$. As mentioned above, we want to decouple the
large Yukawa coupling from $SU(3)_c$ color. Thus we assume $\phi_c$
is charged under the fundamental representation of $SU(3)_c$. The
$\psi_{L/R}^\alpha$ that couple to the Higgs boson carry electroweak
charges, but are color singlets. Both $\phi_c$ and
$\psi_{L/R}^\alpha$ are charged under the new $SU(N)$, and the
massless bound states formed by $(\phi_c\psi_L^\alpha)$ and
$(\phi_c^\dag\psi_R^\alpha)$ are identified as $t_{L/R}^\alpha$.
$\phi_c$ should not be thought as an elementary scalar field;
rather, it is a convenient notation for a product of fermionic
partons \cite{Abbott:1981re, Abbott:1981yg,Dimopoulos:2002bn}.

In Appendix \ref{sec:anomaly}, we give examples with field content and charge assignments consistent with anomaly cancellation.

\section{Fine-tuning and UV Completions}\label{sec:UVcompletion}

The composite nature of the top quark means there are
non-trivial form factors involving the Higgs and the top quark in
the low energy effective theory.  See \cite{Contino:2006nn} for
related discussions of fine-tuning in this context. This will modify
the calculation of the Higgs (mass)$^{2}$. But, as discussed above,
a composite top quark does not eliminate quadratic divergences.
Additional physics is necessary at $\Lambda_\psi$, and fine tuning
is as in Eq. (\ref{eq:QuadHiggs}). The benefit of making the top
composite is that the UV physics at $\Lambda_\psi$ does not
necessarily carry $SU(3)_c$. This can lead to novel phenomenology.

We now sketch two possible UV completions. In the first,  we imagine
the Higgs becomes composite at a (higher) scale,  with ``top
partners" showing up near $\Lambda_{\psi}$. For example, we
may embed this scenario in a Little Higgs-like model
\cite{ArkaniHamed:2001nc,ArkaniHamed:2002qy}, in which the Higgs is
a pseudo-Nambu-Goldstone boson of a strongly coupled sector above
$\Lambda_\psi$. In Little Higgs models, a colored fermionic top
partner emerges from strong dynamics and cuts off the 1-loop
quadratic divergence to Higgs (mass)$^{2}$. Here, the masses of
fermionic partners to $\psi_L$ and $\psi_{R,t}$ set the scale
$\Lambda_{\psi}$. Importantly, these fermionic ``top partners'' partners carry the same quantum numbers as $\psi_L$ and $\psi_{R,t}$, i.e. they are
charged under strong dynamics which confines at $\Lambda_C$ but are
colorless.  Effectively, we have a Little Higgs-like model, but
with $SU(3)_c$ replaced by our new $SU(N)$.  Electroweak divergences
to the Higgs (mass)$^{2}$ may be cutoff as usual in a Little Higgs
model, with new EW resonances (e.g. heavy gauge bosons) appearing at
a couple of TeV.

The fermionic partners of $\psi_L$ and $\psi_{R,t}$ can also combine
with the colored partons present in the composite top.  The result
is  colored composite states, analogous to the top partners of
composite Higgs models. The mass of these states are controlled by
the mass of $SU(3)_c$ neutral partner partons, larger than
$\Lambda_C$. Superficially, the existence of these states makes our
model appear like an ordinary composite Higgs model, since this
means we also have colored top partners whose masses
determine the ultimate fine tuning in the model. However, there is
an important difference: their production at a hadron collider is
dramatically suppressed due to $SU(N)$ hadronization. This
screening effect occurs when the $SU(3)_c$ charge of the composite
particle is only carried by light partons. (If heavy partons in a
composite particle are also charged under $SU(3)_c$, one expects the
colored composite particle has a production comparable to an
elementary colored particle.) The details of this interesting
screening effect will be discussed in Sec. \ref{sec:Screening}.

A second possibility is to UV complete the model supersymmetrically.
In this case, electroweakinos can appear at a low mass scale, cutting off any weak gauge divergences.   The superpartners of
$\psi_L$ and $\psi_{R,t}$, i.e. $\phi_L$ and $\phi_{R,t}$, can be
introduced in order to cut off the remaining quadratic divergences.
These superparticles are uncharged under $SU(3)_c$, thus they do not
have large production rates at a hadron collider. However, just as
above, $\phi_L$ and $\phi_{R,t}$ can combine with colored partons
present in the composite top, forming composite stop states. The
masses of these states will be dominantly determined by the masses of
$\phi_L$ and $\phi_{R,t}$, which are larger than $\Lambda_C$. Again,
the screening effect from $SU(N)$ hadronization dramatically reduces
the production rate of these composite stop states, as discussed in
Sec. \ref{sec:Screening}.

In superymmetric UV completions there is an additional concern: the masses of the new scalars at  $\Lambda_{\psi}$, $\phi_L$ and
$\phi_{R,t}$,  must also be natural.
Since $\phi_L$ and $\phi_{R,t}$ are charged under the strongly
coupled $SU(N)$ gauge group, quantum corrections are mainly from
loops involving $SU(N)$ gauge couplings. One may worry that the
gauge coupling is so strong that the loop expansion is out of
control. This would indicate that the superpartners of the colored
partons (also charged under $SU(N)$) would need to show up at the
same scale, $\Lambda_\psi$ to cut off these divergences.   But in
this case, a superpartner of the colored parton in the composite top,
call it $\tilde {\psi_{c}}$, could combine with the color neutral
partons in the composite top to form another composite stop state,
also charged under $SU(3)_c$. Production of this state would not be
screened, and it would be produced with a cross section similar to
an elementary stop with mass $\Lambda_{\psi}$.

However, there is a subtlety: the $SU(N)$ gauge coupling runs
rapidly near $\Lambda_{C}$ -- it is thus crucial to specify the
energy scale at which the gauge coupling is evaluated when
calculating quantum corrections. We expect the gauge coupling should
be evaluated at $\Lambda_\psi$, i.e. the scale where $\phi_L$ and
$\phi_{R,t}$ appear. (When evaluating the quantum corrections from
the gluino to squarks, we would evaluate $\alpha_{S}$ at the mass
scale of these particles, but not $\Lambda_{QCD}$.) If the running
of $SU(N)$ gauge coupling is fast enough and $\Lambda_\psi$ is not
too close to $\Lambda_C$, the loop expansion is reliable when
calculating the corrections to the masses of $\phi_L$ and
$\phi_{R,t}$. To avoid additional fine tuning, the superpartner of
$SU(3)_c$ colored parton might be as much as 2-loop factors away
from $\Lambda_{\psi}$. Explicitly, we have
\begin{eqnarray}
m^2_{\tilde{\psi_{c}}}\lsim \bigg(\frac{16\pi^2}{Ng^2_{SU(N)}|_{\mu=\Lambda_{\psi}}}\bigg)^2\Lambda_{\psi}^2\label{eq:FineTuningSUSY}.
\end{eqnarray}
If $N g^2_{SU(N)}|_{\mu= \Lambda_{\psi}}$ is not too large, the calculation is under
control, and the mass of the superpartner of the $SU(3)_c$ charged parton present in the top can be parametrically larger than $\Lambda_\psi$.  However, these still might be the first ``stops'' produced at a future hadron collider.

Finally, we  briefly comment on some interesting implications for
the  gluino in our supersymmetric UV completion. In the MSSM, the
gluino generates 1-loop correction to stop mass, which indicates a
2-loop contribution to the Higgs soft mass. The null result of the
gluino search can contribute to the fine tuning. In our scenario,
the gluino mass is related to the superpartner of the $SU(3)_c$
colored parton,  $\tilde \psi_c$ via a loop factor.  However, this
parton only couples to the Higgs indirectly (via $SU(N)$
interactions) and then the Yukawa. Thus it only starts to contribute
to Higgs mass at four-loop level. This likely postpones the
appearance of the gluino.

\section{Screening from $SU(N)$ Hadronization}\label{sec:Screening}

As discussed above, there are heavy
composite particles charged under $SU(3)_c$ in our scenario. These
particles are unavoidable because heavy particles sharing the same
gauge quantum numbers as $\psi_L^\alpha$ and $\psi_{R,t}^\alpha$ are
expected at $\Lambda_\psi$ to truncate the quantum corrections to
Higgs mass. These particles can either be scalars or fermions,
depending on the UV model. Though they themselves are neutral under
$SU(3)_c$, they can combine with the colored constituents $\phi_c$
that appear in the composite top and form heavy composite colored
particles. In this section, we will show that the production of
these heavy colored composite particles can be dramatically
suppressed due to $SU(N)$ hadronization. Hadron collider
constraints on these composite top partners can thus be effectively
removed or delayed.

For concreteness, consider an example where the SUSY partners of
$\psi_{L/R}^\alpha$, i.e. $\phi_{L/R}$, are introduced at
$\Lambda_\psi$, a little higher than $\Lambda_{C}$.\footnote{Similar
arguments apply to fermionic top partners if the quadratic
divergence is ultimately softened via a Little Higgs-like model.}
These particles cancel the corrections to the Higgs mass from the
$y_\psi H\psi_L\psi_R$ Yukawa coupling present above $\lambdac$.
Since $\phi_{L/R}$ are singlets under $SU(3)_c$, they are directly
produced solely by electroweak interactions. The direct
production rate is therefore much lower than the analogous particles
responsible for cutting of divergences in the Minimal Supersymmetric
Standard Model (MSSM), the stop.  However, $\phi_{L/R}$ carry the
same $SU(N)$ charges as $\psi_{L/R}^\alpha$. They can combine with
$\phi_c$ and form composite stop-like resonances with mass dominated by the mass of $\phi_{L/R}$.

Although the composite stops states carry $SU(3)_c$ charge, QCD
processes can only produce these colored heavy resonances indirectly
-- through $SU(N)$ hadronization.  Because the mass of this particle
is greater than $\Lambda_{C}$, one must first pair produce the
$\phi_c$ via QCD. Hadronization of the $SU(N)$ gauge group may
in principle produce the heavy parton $\phi_{L/R}$ which could combine with
$\phi_c$ and form stop-like states.  However, the production of
$\phi_{L/R}$ from $SU(N)$ hadronization is highly suppressed if its
mass is higher than $\Lambda_{C}$. To estimate the probability of
$\phi_{L/R}$ production, we rely on the string fragmentation
approach \cite{Andersson:1983ia}. Quarks of different flavors are
produced through quantum tunneling with probabilities as:
\begin{eqnarray}
 \label{Lund}
P_{q_i} \propto e^{-\pi m_i^2/\kappa}.
\end{eqnarray}
Here $m_i$ is the quark mass and $\kappa$ is the string tension in
QCD, $\kappa\simeq \Lambda_{QCD}^2\simeq 0.2 \;\textrm{GeV}^2$. Even
a modest hierarchy between $\Lambda_{QCD}$ and $m_{q}$ has a
dramatic effect: the production of charm quark ($m_{c} \simeq$ 1.29
GeV) via this mechanism is already eleven orders of magnitude
smaller than the production of strange quark. We expect a similar
suppression when an energetic $\phi_c$ is produced and hadronizes
under the $SU(N)$ gauge group. It should go almost exclusively to
composite particles formed by light partons, i.e. third
generation quarks.\footnote{The confinement of $SU(N)$ might produce
other light composite particles below $\Lambda_C$, other than top
and bottom quarks. The spectrum and charges of these particles are
model dependent , and could have important consequences for the
collider phenomenology of this scenario; see discussion in Sec.~\ref{sec:collider}. } This provides an
interesting way to suppress the production of the composite stop
even though the $\phi_{L/R}$ are only a little bit heavier than
$\Lambda_{SU(N)}$.

Finally, we re-emphasize that this screening effect only applies to
the cases where the $SU(3)_c$ colored parton is light, with mass of
the composite particle controlled by the uncolored heavy parton with
mass beyond $\Lambda_C$. If the heavy parton of the composite
particle is colored, these heavy partons can be directly produced in
hadron colliders. Naturalness considerations do not require these
particles as light, however, see discussion around
Eq.~(\ref{eq:FineTuningSUSY}).

\section{Phenomenological Constraints}\label{sec:constraints}

$\lambdac$ is crucial to determining the amount of fine-tuning in
this scenario. We now discuss how low $\lambdac$ can be, consistent
with existing experiments. We consider three classes of constraints:
electroweak precision tests (EWPT), direct collider bounds, and
flavor physics.  Constraints actually exist on two different but
related scales.  There are direct constraints on both the
compositeness scale $\lambdac$ and on the scale $f \equiv
\lambdac/g_{\ast}$, where $g_{\ast}$ is a typical (strong) coupling
among heavy resonances, such as baryonic and mesonic states,
participating in the strong dynamics. In NDA \cite{NDA1, NDA2},
$\lambdac = 4 \pi f$. We will track constraints on these two scales
separately, to allow for violation of the naive NDA relation.

We estimate effects induced by strong dynamics by integrating out
heavy resonances whose masses are naturally ${\mathcal O}
(\Lambda_C)$. The elementary Higgs boson interacts with the strongly
coupled sector via tree-level, perturbative Yukawa couplings with
the top and bottom quarks and their excited states.

\subsection{Electroweak precision tests}
\label{sec:ewpt}

The effects of this strong dynamics on precision EW observables can
be characterized in terms of effective operators. A similar overview
of EWPT in theories of top compositeness can be found in
\cite{Georgi:1994ha}. In this section, we lay out the classes of
effective operators that are generated and discuss the constraints
on each. Here, we focus on flavor diagonal operators, postponing the
important issue of flavor changing neutral currents (FCNCs) until Section \ref{sec:flavor}.

Operators coupling third generation fermions to bosons include:
\begin{eqnarray}
{\mathcal O}_{1} &=&i c_1  (H^{\dagger} D^{\mu} H) (\bar{\Psi} \gamma_{\mu} \Psi) + h.c. \label{eq:Op1}, \\
{\mathcal O}_{2} &=& i g c_2  \, \bar{\Psi} W_{\mu\nu}D^\mu
\gamma^\nu \Psi \label{eq:Op2},
\end{eqnarray}
where $\Psi$ represents a composite (third generation) fermion,
$c_1$ and $c_2$ represent coefficients of dimension [mass]$^{-2}$.
Other operators such as $i \bar\Psi D^2 \slashed{D}\Psi$ can be
related to the above operators (along with others not well
constrained) by using equations of motion \cite{Buchmuller:1985jz}.
The operator of Eq.~(\ref{eq:Op1}) is most strongly constrained by
precision measurements of couplings of the $Z$-boson to $b$-quarks.
The operator of Eq.~(\ref{eq:Op2}) gives rise to an imaginary
Feynman rule, and thus does not interfere with the SM, except
suppressed by the $Z$-width. Constraints on it are substantially
weaker.

To determine the constraints from the operator of Eq.~(\ref{eq:Op1}), we begin by writing the coupling of the $Z$ to $b$ quarks as:
\begin{eqnarray}
 \label{Zbb}
L\supset \frac{g}{c_W}Z_\mu(g^L_{b}\bar b_L \gamma^\mu b_L+g^R_{b}
\bar b_R\gamma^\mu b_R).
\end{eqnarray}
Here $c_W\equiv \cos{\theta_W}$ and $g$ is the $SU(2)_L$
gauge coupling. We define the new physics contributions to $Z\bar b b$
couplings through
\begin{eqnarray}
 \label{ZbbNP}
g^L_{b}=g_{b}^{L,SM}+\delta g^L_{b},\ \ g^R_{b}= g_{b}^{R,SM}+\delta
g^R_{b},
\end{eqnarray}
where tree level SM values are $g_{b,0}^{L,SM}=-1/2+s_W^2/3\simeq
-0.42$ and $g_{b,0}^{R,SM}=s_W^2/3\simeq 0.077$. The dominant
constraint is expected to arise from measurement of $R_{b}$.  This
most strongly constrains operators in Eq.~(\ref{eq:Op1}) with $\Psi
= b_{L}$, effectively a contribution to $\delta g^L_b$.  Following
\cite{Gori:2015nqa}, we write
\begin{equation}
\delta R_{b}^{0} \approx -0.78\delta g^L_{b} + 0.14 \delta g^R_{b}.
\end{equation}
Using the 2$\sigma$ experimental uncertainty $\delta R_{b}^0 =
0.0013$, we find $\delta g^L_{b} <1.7 \times 10^{-3} $.  A larger
value could be accommodated if a positive shift to $g_R^b$ is
present, though eventually, this would conflict with measurements of
$A_{FB}^b$.\footnote{At present $A_{FB}^{0}$ is slightly
discrepant from the SM prediction, so a small positive contribution
to $g_{R}^{b}$ is favored.}

\begin{figure}
\centering
\vspace{-1.25in}
%\subfloat[Loop contribution, utilizing strong dynamics to couple four third generation fermions together.]{
\includegraphics[width=0.5\textwidth]{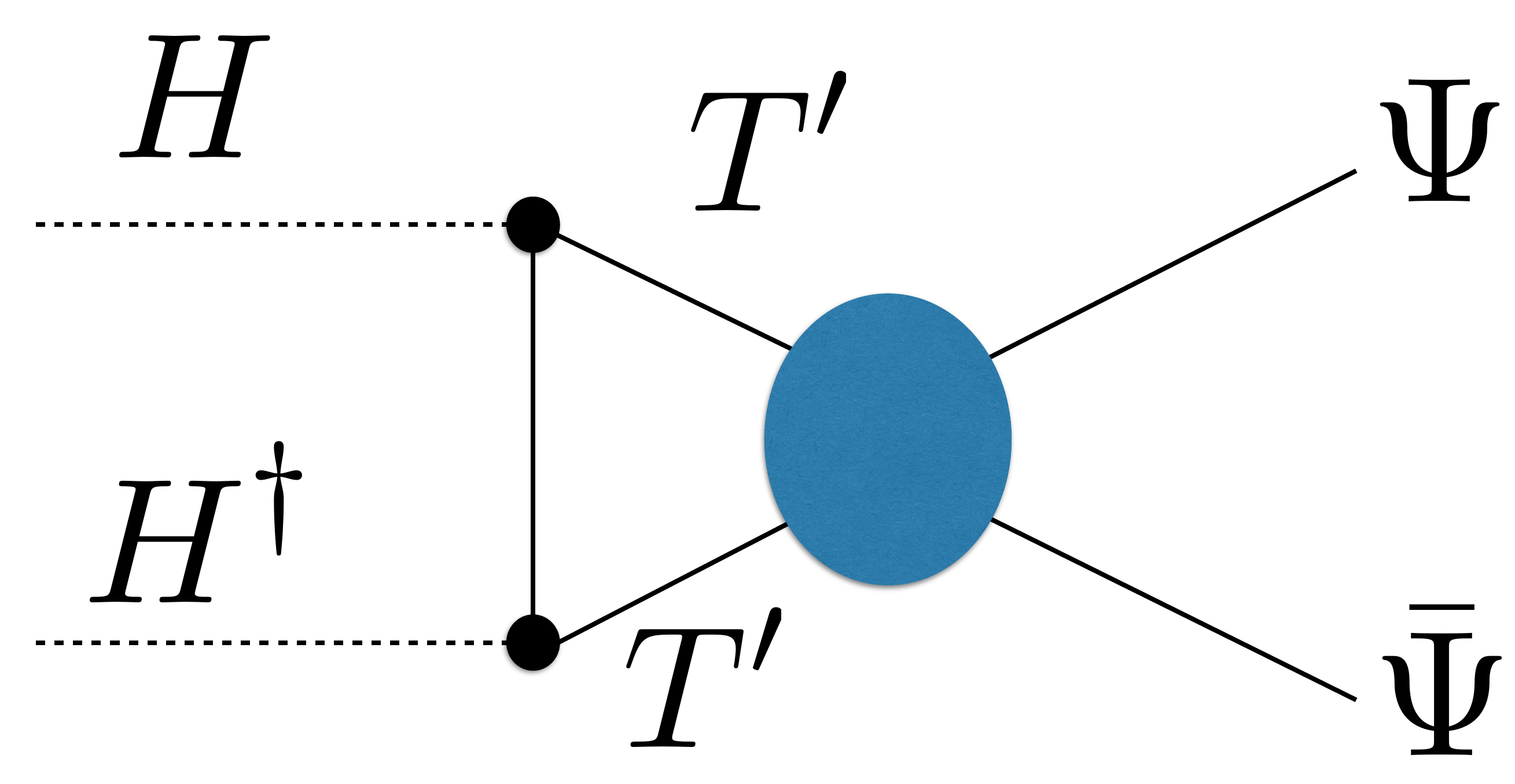}
\caption{A loop induced contribution to $Z-b-b$ coupling, utilizing strong dynamics to couple heavy top partners to SM third generation fermions.}
\label{fig:bbOps}
%\vspace{-.1in}
%\subfloat[Tree contribution utilizing a form factor at the Higgs bottom quark vertex.]
%{
%\includegraphics[width=0.3\textwidth]{BlobFF-crop.pdf}
%\label{fig:bbTree}
%}
%\caption{Contributions to the $(H^{\dagger} D_{\mu} H)(\Psi \gamma^{\mu} \Psi)$ operator.}
%\label{fig:bbOps}
\end{figure}

The operator of Eq.~(\ref{eq:Op1}) can be generated after
integrating out strong dynamics, see Fig.~\ref{fig:bbOps}. Here we
use $T^{\prime}$ to stand for heavy resonances that emerge from
strong dynamics which couple to the Higgs boson (excited states of
the top may be an example). Such resonances can be converted to $b$
quarks via the strong dynamics (e.g. via the exchange of a $\rho'$
of the strong dynamics). The effect of the $\rho'$ exchange is a
4-fermion operator suppressed by $f$. The details of these operators
will be discussed in Sec. \ref{sec:4fermi}. Since the Higgs boson is
assumed to be an elementary particle, it only couples to particular
partons charged under the strong dynamics. The largest coupling
between Higgs and partons is the $y_{\psi}$ which induces $y_t$
after confinement. Let us take the excited top state $T'$ (whose
collider phenomenology will be discussed in Sec. \ref{sec:collider})
as an example, the coupling between Higgs and $T'$ is expected to be
similar to $y_t$. The result is a contribution going parametrically
as:
\begin{equation}
c_{1}^{loop} = \frac{\eta_{b}^{loop} N_c y_{t}^2}{16 \pi^2 f^2} .
\end{equation}
Here $\eta_{b}^{loop}$ is a (presumably ${\mathcal O}(1)$) unknown
number.  This constrains $f \gsim \sqrt{\eta_b^{loop}} \, 410$ GeV.
Note, additional corrections to the $Z\bar b b$ coupling can be induced
through operators such as $(H Q_L)^\dag \slashed{D} (H Q_L)$. Such
operators can be generated by integrating out heavy resonances of
$b_R$. However, these contributions are proportional to $y_b^2$ and
thus negligible.

In theories of compositeness/strong dynamics, the oblique parameters $S$ and $T$ \cite{Peskin:1991sw,Peskin:1990zt}
often provide an important constraint. However, in our theory, where the Higgs remains elementary, these constraints are weakened.

The relevant operators can be written as
\cite{Grinstein:1991cd,Han:2004az}
\begin{eqnarray}\label{STpara}
{\mathcal O}_{S} &=& c_{S}  (H^{\dagger} \tau^{a} H) W^{a}_{\mu \nu} B^{\mu \nu},\\
{\mathcal O}_{T} &=& c_{T} |H^{\dagger} D_{\mu} H|^2,
\end{eqnarray}
with
\begin{eqnarray}
c_{S} &=& \frac{1}{4 s c} \frac{\alpha}{v^2} S,\\
c_{T} &=& -2 \frac{\alpha}{v^{2}} T.
\end{eqnarray}
In this case, there are constraints both directly on the
compositeness scale $\Lambda_C$ as well as on the scale $f$.  The
contribution directly constraining $\Lambda_C$ arises from
integrating out  heavy states such as the $T'$ (there is a box
diagram).  The second contribution comes from coupling Higgs bosons
to the $T'$, and then using the four-fermion verxtex involving four
$T'$'s suppressed by $f^{2}$.  For $c_T$, we have contributions
\begin{equation}
c_{T} = \frac{\eta_{T}^{\Lambda} N_{c} y_t^4}{16 \pi^2 \lambdac^2} , \, \, \frac{  \eta_{T}^{f} N_{c}^2 y_t^4}{(16 \pi^2)^2 f^2}.
\end{equation}
These contributions are equal for $\lambdac = 4 \pi f$ as assumed in NDA.  For the $S$ parameter, we have:
\begin{equation}
c_{S} = \frac{\eta_{S}^{\Lambda} N_{c} y_t^2 g g^{\prime}}{16 \pi^2 \lambdac^2} , \, \, \frac{\eta_{S}^{f} N_{c}^2 y_t^2 g g^{\prime}}{(16 \pi^2)^2 f^2},
\end{equation}
which are again equal under the NDA assumption. Imposing approximate
2$\sigma$ bounds $S < 0.16$ and $T < 0.22$, we find $S$ implies
$\lambdac > \sqrt{\eta_{S}^{\Lambda}}$ 1200 GeV, while $T$ implies
$\lambdac > \sqrt{\eta_{T}^{\Lambda}}$ 610 GeV. For the oblique
parameters, we expect the bounds directly dependent on $\lambdac$ to
be the strongest constraints, as  $\lambdac < 4 \pi f$.

\subsection{4-Fermi operators}
\label{sec:4fermi}
We now turn to the four-fermion interaction term, i.e.
\begin{eqnarray}
\label{eqn:Op4fermi}
{\mathcal O}_{4f}=c_{4f} (\bar{\Psi}\gamma^\mu\Psi)(\bar\Psi\gamma_\mu\Psi).  \label{eq:Op3}
\end{eqnarray}
We expect $c_{4f} = \eta_{4f}/f^{2}$, with  $\eta_{4f}$ a presumably
order one number.  Such interactions are introduced, e.g. by the
exchange of resonances of the strong dynamics.

It was proposed in \cite{Lillie:2007hd,Kumar:2009vs} that the search
in four top channels at the LHC could impose strong constraints on
composite top scenarios. See also
\cite{Fabbrichesi:2013bca,Englert:2014oea,Pomarol:2008bh}. Such a
study has been carried out by ATLAS at 13 TeV with 3.2 fb$^{-1}$
data \cite{ATLAS4Top}. $c_{4f}$ is constrained at the 95\% confidence level as
\begin{eqnarray}
 \label{4top}
c_{4f} <\frac{5}{\rm{TeV}^2},
\end{eqnarray}
which implies $f> \sqrt{\eta_{4f}}$ 450 GeV. Depending on the
precise value  of $\eta_{4f}$, this constraint can either be weaker
or stronger than the indirect constraints arising from $Z
\rightarrow b \bar{b}$ discussed in the previous section ($f>
\sqrt{\eta_{b}^{loop}}$ 410 GeV).  Importantly, we expect the bounds
on this operator to improve dramatically with new LHC data.

It may be possible that the rank of strongly coupled group, i.e.
$N$, or the representations of partons under the new strong dynamics
may give an argument for an effective suppression of $c_{4f}$.  To
flesh out how this might occur, we first review an example from QCD,
and then contrast how things would differ with more exotic
representations.  We build our way up to derive the $N$ counting for
a baryon-baryon interaction mediated by pion exchange. The interaction between a pion and partonic quarks scales as
$\frac{1}{\sqrt{N}}$:  the two point correlation function between
two pions has an $N$ enhancement, due to a closed color loop, and
$\frac{1}{\sqrt{N}}$ appears in the pion-quark interaction vertex
after proper normalization. The interaction between a baryon (formed
by $N$ quarks in fundamental representation) and a pion scales as
$\sqrt{N}$: baryons carry $N$ colored lines, allowing $N$ ways to
insert the pion-quark vertex into a baryon. When combined with the
factor of $\frac{1}{\sqrt{N}}$ from the pion-quark vertex, the
baryon-pion interaction scales as $\sqrt{N}$. This implies that the
baryon-baryon interaction mediated by pion exchange scales as $N^0$
because there is only one way to insert the pion into a second
baryon after the color has been fixed via the choice on the first
baryon.

This discussion relies on the assumption that the baryon is formed by $N$ quarks
which are in the fundamental representation of $SU(N)$.
If the partons transform in a more complicated representation, then the $N$-scaling differs.  For example, if the new strong dynamics is given by $SU(6)$, then the top quark could conceivably be comprised of
three quarks, $\psi_{1,2,3}$, transforming as $\textbf{6}$,
$\textbf{15}$, $\textbf{20}$ representations, respectively. Although color-singlet mesons
 still take the form $M_i\sim\bar\psi_i\psi_i$,  the meson-parton vertex now scales as
$(\frac{1}{\sqrt{N}})^i$. This is because the $i^{\rm{th}}$ parton
carries $i$ color indices and there are $i$ closed color lines when
computing the two point correlation functions of these mesons, There
is no compensating factor, as there is only one way to insert a
meson to a baryon when computing the interaction mediated by these
mesons: the meson-parton vertex vanishes if the meson is composed by
a different species of parton. So, it seems plausible that
interactions among baryons may be suppressed if the partons
transform in complicated representations.  Of course, whether that
obtains is a model-dependent question.

\subsection{Flavor/CP constraints}
\label{sec:flavor}
\subsubsection{Flavor constraints}

Flavor is a concern in models where the third generation quark
is composite. As discussed in \cite{Georgi:1994ha}, once the
fermions in Eq. (\ref{eq:Op1}), (\ref{eq:Op2}) and (\ref{eq:Op3})
are rotated to the mass basis, severe flavor problems may result.
Let us explicitly illustrate how flavor changing processes are
introduced. The mixing between the first two generations and the
third generation quarks can be characterized by the following
approximate rotations, which take quarks from the interaction ($I$)
basis to the mass ($M$) basis:
\begin{eqnarray}
 \label{RotationBasis}
t_{L/R,I}&\approx& \theta_{L/R}^u u_{L/R,M}+\theta_{L/R}^c
c_{L/R,M}+t_{L/R,M},\nonumber\\
b_{L/R,I}&\approx & \theta_{L/R}^d d_{L/R,M}+\theta_{L/R}^s
s_{L/R,M}+b_{L/R,M}.
\end{eqnarray}
In our scenario, only operators involving the third generation are
directly generated by the strong dynamics. This argues the operators
Eq. (\ref{eq:Op1}), (\ref{eq:Op2}) and (\ref{eq:Op3}) naturally only
involve third generation partons in the interaction basis.

Operators connecting the first two generations with the partons
forming the third generation may be generated at a higher scale.
Indeed, the observed CKM matrix indicates that the mixing between the third
generation and the first two generations is small (but not zero). Non-renormalizable operators above confinement scale can be responsible for this mixing:
\begin{eqnarray}
 \label{CKMGenerate}
\frac{1}{\Lambda_{\textrm{mixing}}^{3(m-1)/2}}H \psi_s^m \psi_{1,2},
\end{eqnarray}
where $\psi_s$ generically labels partons of the third generation
quarks in strong dynamics, and we assumed there are $m$ partons
combined together to form a third generation quark. $\psi_{1,2}$
stands for the first two generation elementary quarks.
$\Lambda_{\textrm{mixing}}$ is a scale higher than $\Lambda_C$ where the mixing is generated.
Passing through the confinement scale, $\psi_s^m$ becomes a single
composite fermion in the IR (a third generation quark), and the high dimension operators become effective marginal operators,
\begin{eqnarray}
 \label{CKMGenerate2}
\frac{1}{\Lambda_{\textrm{mixing}}^{3(m-1)/2}}H \psi_s^m
\psi_{1,2}\to
\bigg(\frac{\Lambda_{C}}{\Lambda_{\textrm{mixing}}}\bigg)^{3(m-1)/2}H
\psi_3 \psi_{1,2}.
\end{eqnarray}
Such operators generate the mixing
between the third generation quarks and the first two
generations.\footnote{In principle, there are higher dimension
operators at scale $\Lambda_{\textrm{mixing}}$ that generate flavor
changing operators not confined to the third generation, e.g.
$\frac{1}{\Lambda_{\textrm{mixing}}^{3m-1}} \psi_s^m \psi_s^m
\psi_{1,2}\psi_{1,2}$.  Comparing to Eq.~(\ref{CKMGenerate2}), and
relating the effective suppression to CKM suppression we expect
these operators to be at least as suppressed as those described
below, and we do not discuss them further. An approximate flavor
symmetry may be useful in suppressing some of these operators in the
down-type sector.}  This mixing is dangerous when
combined with the operators generated by strong dynamics that we now
discuss.

There are four independent classes of
operators relevant to flavor changing processes:
\begin{eqnarray}
 \label{FlavorOp}
\mathcal{O}_{F1}&=&\frac{e^2}{\Lambda_C^2}(\bar\Psi_{L/R}
\gamma^\mu\Psi_{L/R})( \bar l \gamma^\mu l),\nonumber\\
\mathcal{O}_{F2}&=&e \frac{m_\Psi}{\Lambda_C^2}F_{\mu\nu}(\bar\Psi_{L/R} \sigma^{\mu\nu}\Psi_{R/L}),\nonumber\\
\mathcal{O}_{F3}&=&\frac{1}{f^2}(\bar\Psi\gamma^\mu\Psi)(\bar\Psi\gamma_\mu\Psi), \nonumber \\
\mathcal{O}_{F4} &=& \frac{1}{16 \pi^{2} f^{2}}(H^{\dagger} D_{\mu} H) (\bar{\Psi} \gamma^{\mu} \Psi).
\end{eqnarray}
Prior to mixing, the $\Psi$ are third generation fermions. The first
two classes of operators induce processes with $\Delta F=1$
such as $b\to s \gamma$ or $b\to s \ell^+\ell^-$. The first operator
in Eq. (\ref{FlavorOp}) is related to the operator with photon field
strength by equations of motion, i.e.
$e\frac{i}{\Lambda^2}F_{\mu\nu}\bar\Psi_{L/R}
D^\mu\gamma^\nu\Psi_{L/R}$. The last operator is analogous to the
operator discussed in the context of  $Z \rightarrow b \bar{b}$, but
here (after mixing) we allow for flavor dependence on the coupling.
For most processes, under the assumption that $4 \pi f > \Lambda_C$
the effect of this operator will be subdominant to the effects of
the other operators.

The mixing angles in Eq. (\ref{RotationBasis})
are not arbitrary, since they are related to elements in the CKM
matrix. Under the approximation that all $\theta$'s are small,
these mixing parameters can be related to the elements in CKM matrix
at leading order as
\begin{eqnarray}
 \label{CKM}
|\theta_L^{d*}+\theta_L^u|&\simeq&|V_{ub}|,\nonumber\\
|\theta_L^{s*}+\theta_L^c|&\simeq&|V_{cb}|,
\end{eqnarray}
where $|V_{cb}|\simeq 0.04$ and $|V_{ub}|\simeq 0.004$. The mixing
angles of left-handed quarks for both up and down-type quarks cannot
be simultaneously be taken arbitrarily small.  As we will see,
the constraints arising from FCNC involving $b$ quarks are
particularly strong. It is therefore advantageous to induce the
quark mixing mainly via the up-type Yukawas
($\theta_{L}^d\simeq\theta_{R}^d\simeq 0$). Making this choice, the
dominant constraints come from D-meson mixing.

Before discussing this in detail, let us first emphasize the danger
of allowing appreciable mixing in the down sector.
The constraints on B-meson oscillations have been studied
in \cite{Calibbi:2012at, Altmannshofer:2014rta}, and we will reinterpret their results accordingly. We first consider the operators with all left-handed
fermions, i.e.
$\frac{1}{f^2}(\bar\Psi_L\gamma^\mu\Psi_L)(\bar\Psi_L\gamma_\mu\Psi_L)$.
The constraints
from $B$-meson oscillation can be derived as
\begin{eqnarray}
 \label{BOsc1}
f&\gsim & 2.3 \, \bigl(\frac{\theta_{L}^d}{0.004}\bigr)\ \textrm{TeV}\ \ \ \ \ B_d\ \textrm{oscillation,}\nonumber\\
f&\gsim & 10  \, \bigl(\frac{\theta_{L}^s}{0.04}\bigr)\ \textrm{TeV}\ \ \ \ \ B_s\
\textrm{oscillation.}
\end{eqnarray}
We have normalized the angles to the relevant CKM mixings.  If left-handed and right-handed mixing angles are comparable to each
other, the strongest constraints are imposed to mixed chirality
operators, i.e. $\frac{1}{f^2}(\bar\Psi_L\Psi_R)(\bar\Psi_L\Psi_R)$:
\begin{eqnarray}
 \label{BOsc2}
f&\gsim & 6.2 \sqrt{\frac{\theta_{L}^d}{0.004}}\sqrt{\frac{\theta_{R}^d}{0.004}}\ \textrm{TeV} \ \ \ \ \ B_d\ \textrm{oscillation,}\nonumber\\
f&\gsim & 26
\sqrt{\frac{\theta_{L}^s}{0.04}}\sqrt{\frac{\theta_{R}^s}{0.04}}\
\textrm{TeV}\ \ \ \ \ B_s\ \textrm{oscillation.}
\end{eqnarray}
The normalization of the $\theta_R$ is somewhat arbitrary, as it is not directly related to a CKM angle.  Here, we have considered the constraint on the real part of the operator; the constraint on the imaginary part is modestly ($\sim$ factor 2) stronger.

In the Appendix, we briefly review other bounds coming from down-type
mixing, but motivated by the severity of the above bounds, we will suppose that
quark mixing is generated via the up-type quarks.

$\Delta C=2$ processes, i.e. D-meson oscillation, are induced from
$\mathcal{O}_{F3}$ in Eq. (\ref{FlavorOp}):  the effective operators
are proportional to $(\theta_{L/R}^c\theta_{L/R}^u)^2$. While the
operator in Eq. (\ref{FlavorOp}) containing all left-handed top
quarks is necessarily real, following the rotation to the mass
basis, a contribution to CP-violation in the charm sector appears.
This imposes a particularly stringent constraint, as the
Standard Model contribution is expected to be very small. Indeed,
when all quark mixing arises from the left-handed up quarks, not
only are $\theta_L^{u/c}$ are responsible for generation of
$|V_{ub}|$ and $|V_{cb}|$ in CKM matrix, the phases of
$\theta_L^{u/c}$ are related to the physical phase in CKM matrix.

Let us discuss this constraint in some detail:  CP-violation in the
D-meson system has been constrained via a variety of final states.
Particularly relevant are: $D^0\to K^+\pi^-$, $D^0\to K^+ K^-$ and
$D^0\to \pi^+ \pi^-$. The strongest constraints typically come from
$D^0\to K^+\pi^-$, but this statement depends on whether there is
direct CP violation (CPV) in doubly Cabbibo suppressed ($c\to d u
\bar s$) decays. If present, this additional free parameter can
``soak-up'' CPV in the mixing in $D^0\to K^+\pi^-$, via a
cancellation.  In our scenario, however, we expect this direct CPV
to be small, so the stronger constraints apply \cite{HFAG}.\footnote{Mixing
between $t_L$ and $c_L$ induces a tiny direct CPV
starting with $t\to d u \bar s$ and mixing top with charm by
$\theta_L^c$. However, this is very small, proportional to
$V_{td}V_{us}^*\theta_L^c$. The CPV induced is smaller than $0.1\%$,
much smaller than the error bar on CPV in the $D^0\to K^+\pi^-$
process, i.e. $1\sim10\%$.}

Before presenting detailed numbers, we show the CP violation in our
scenario is invariant under the reparameterization of the CKM matrix.
As an example, consider the $D^0\to \pi^+ \pi^-$ channel. Since we
have set the mixings in down-type sector to vanish, we have
$\theta_L^u=V_{ub}^*$ and $\theta_L^c=V_{cb}^*$. Staring with
$\mathcal {O}_{F3}$ in the interaction basis, we rotate the quarks
into mass basis, and find
\begin{eqnarray}
 \label{cucu}
c_{\mathcal {O}_{F3}}(V_{ub}^* V_{cb}) (\bar c_L\gamma^\mu u_L)(
\bar c_L\gamma_\mu u_L).
\end{eqnarray}
This operator induces mixing between $D^0$ and $\bar D^0$. Including
the D-meson decay in the $\pi\pi$ channel we find CPV proportional
to $\textrm{Im}[(V_{ub}^* V_{cb})^2 (V_{cd}^* V_{ud})^2]$, which is
CKM reparameterization invariant \cite{Jarlskog:1985ht}.  For
simplicity, we choose the standard parametrization where the
CP-violating phase is primarily moved to $V_{td}$ and $V_{ub}$. From
\cite{HFAG}, we choose CP violating parameters within the 2$\sigma$
allowed region, i.e. $x_{12}= 0.4\%$ and $\phi_{12}=
5^\circ$.\footnote{Here $x_{12}\equiv 2|M_{12}|/\Gamma$ and
$\phi_{12}\equiv \textrm{arg}(M_{12}/\Gamma_{12})$, where $M_{12}$
and $\Gamma_{12}$ are $\bar{M^0}-M^0$ transition amplitudes, i.e.
$\langle M^0|H|\bar{M^0}\rangle=M_{12}-\frac{i}{2}\Gamma_{12}$.}
This translates to
\begin{eqnarray}
 \label{CPVconstraint}
f\gsim \sqrt{\eta_{4f}} \, 810 \, \textrm{GeV}.
\end{eqnarray}
We have included the RG running on the operator following \cite{Golowich:2009ii}.  This is the strongest constraint on $f$ from indirect measurements.
A $40\%$ tuning (cancellation) among the four-fermion
operators (perhaps via a tiny mixing amongst the right-handed
up quarks) would reduce this constraint below the
direct constraint on $\mathcal {O}_{F3}$ from 4-top production at the LHC,  $f> 450$ GeV.

The mixing in Eq. (\ref{RotationBasis}) also induces $\Delta C=1$
rare D-meson decays, such as $D^+\to \pi^+ +\mu^+ +\mu^-$ and $D \to
\mu^{+} \mu^{-}$.  The effective operators after rotation are
proportional to $(\theta_{L/R}^c\theta_{L/R}^u)$. However, The presence of large long-distance contributions to these decays weaken
these constraints, and following \cite{Fajfer:2015mia}, we find
these constraints are subdominant to those derived from
mixing.

We briefly note the potential importance of a dimension 6 operator beyond those considered in Eq. (\ref{FlavorOp}) that is relevant independent of how quark mixing is introduced:
\begin{eqnarray}
 \label{Wtb}
i (H^{c\dag}D^\mu H) (\bar t_R \gamma_\mu b_R).
\end{eqnarray}
This operator can induce a coupling between W boson and right-handed top/bottom quarks and has been studied in the context of composite Higgs
models in \cite{Vignaroli:2012si}. Because the SM contribution to $b\to s\gamma$ suffers from helicity suppression, i.e. proportional to $m_b$,
coming from the mass insertion on the external leg. If the operator
in Eq. (\ref{Wtb}) is present, the mass insertion is not required
anymore. Further, the SM contribution is loop suppressed, so even a
fairly small coefficient in front of this operator may induce a
too-large contribution to $b\to s\gamma$.  Fortunately, in our scenario the
coefficient is expected to be suppressed by $y_b$. Consider the limit where $y_b$ is
set to zero: one may assign a conserved quantum number to $b_R$.
If this is respected by strong dynamics, the bottom Yukawa
coupling is the only interaction which violates this $b_R$ number.
Thus the coefficient in Eq. (\ref{Wtb}) has to be proportional to
$y_b$. Similarly, $y_t$ should also appear in the coefficient, and
the operator can be written as
\begin{eqnarray}
 \label{Wtb2}
i\frac{y_t y_b \eta_{Wtb}}{\Lambda_C^2}(H^{c\dag}D^\mu H) (\bar t_R \gamma_\mu b_R).
\end{eqnarray}
Applying the constraints from $b\to s\gamma$, we have
\begin{eqnarray}
 \label{Wtb3}
\Lambda_C\gsim \sqrt{\eta_{Wtb}} \, 930 \,  \textrm{GeV}.
\end{eqnarray}

At last, the electric dipole operator (EDM) of top quarks can also
be constrained since it contributes to neutron EDM at low energy
\cite{Kamenik:2011dk}. Prior to electroweak symmetry breaking, these
operators can be written as dimension 6 operators:
\begin{eqnarray}
 \label{EDM}
\frac{c_{EDM} g_i}{\lambdac^2} \bar Q H \sigma^{\mu\nu} U F^i_{\mu\nu},
\end{eqnarray}
where $i$ stands for SM gauge groups. $F^i_{\mu\nu}$ is the field
strength of the $i$th group and its gauge index is contracted with
the corresponding generator which is implicitly included. The bound
on the neutron EDM requires $\textrm{Im}(\lambdac/\sqrt{c_{EDM}}) >
4.7\ \textrm{TeV}$. Thus if $\lambdac$ is 1$\sim$2 TeV, for $c_{EDM}
= 1$, one needs a mild suppression of CP violating phases in the
strongly coupled sector $\phi_{CP} \lsim 0.1$.

\subsection{Mixing between elementary and composite degrees of freedom}\label{PhenoMixing}

The discussion on phenomenological constraints so far ignores the
possible mixing between elementary degrees of freedom and composite
degrees of freedom. In this section, we present a general analysis
and show that the mixing generically does not introduce stronger
constraints.

First, $\psi_L$ and $\psi_R$ can also form a composite scalar,
$\Phi_H$, with quantum numbers identical to those of the SM Higgs
boson. In the present set-up $\Phi_H$ is not a pseudo-Nambu
Goldstone boson, so its (unprotected) mass is expected to be around
confinement scale: $m_{\Phi_H}\sim \Lambda_C$. The elementary SM
Higgs can mix with $\Phi_H$. While the mixing cannot be calculated in a
precise way, we will argue that expected effects induced by such mixing are
comparable or smaller than those enumerated from direct couplings to the
elementary Higgs boson.

Although we cannot rule out the possibility of a ``tree-level'' bare
mixing directly present between the elementary Higgs boson and the
composite state, we try to estimate its natural size by calculating
the loop contribution to this mixing arising from integrating out
heavy resonances. The loop contribution can be estimated by
integrating out heavy resonance modes from strong dynamics which
couple to both Higgs and $\Phi_H$. For example, integrating out the
excited state of top quark induces a mixing term in Lagrangian as
$\mathscr{L}\sim \frac{N_c y_t g_*}{16\pi^2\sqrt{N}}\Lambda_C^2
H\Phi_H $. \footnote{Note $\sqrt{N}$ in the denominator appears due
to the large-N scaling on meson-baryon coupling. Unlike the
conventional baryon-meson coupling scaling, here we assume that
there is only one consistent way to insert $\Phi$ to partons in
$T'$. A more detailed discussion can be found at the end of Sec.
\ref{sec:4fermi}.  } For a composite state with mass $\Lambda_C$,
the mixing angle be estimated to be $\sim\frac{N_c y_t g_*}{16\pi^2
\sqrt{N}}$. As studied in previous sections, $g_*$ is expected to be
smaller or comparable to $4\pi$. Thus, with a reasonable choice of
$N$, the mixing can be easily smaller than $10\%$ and modifications
of Higgs boson properties due to this mixing should be consistent
with present measurements.

Higher dimension operators containing $\Phi_H$ suffer a smaller
effective suppression scale than those operators containing the $H$
directly as $\Phi_H$ has strong couplings to other composite states.
These operators containing $\Phi_H$ will generate operators with $H$
once we account for the above mixing.  One might worry that this
might induce the dominant contribution to precision electroweak observables, but the
mixing compensates for the would-be smaller suppression scale, and
these induced operators are expected to be subdominant. To see this,
consider operators differ by one power of $H$ and
$\Phi_H$. They can be written as $\frac{y_t H}{\Lambda_C}\mathcal
{O}$ and $\frac{g_* \Phi_H}{\sqrt{N}\Lambda_C}\mathcal {O}$.
Performing the rotation from $\Phi_H$ to $H$, the second operator
can induces  $\frac{ g_*^2 y_t N_C}{ 16\pi^2
N\Lambda_C}H\mathcal {O}$. So, effects induced from $\frac{g_*
\Phi_H}{\Lambda_C}\mathcal {O}$ are comparable or smaller than those
from $\frac{y_t H}{\Lambda_C}\mathcal {O}$, since $g_*<4\pi$,
especially when $N$ is large. This can be generalized to operators
involving an arbitrary number of Higgs fields.

Furthermore, our lighter Higgs boson is mainly elementary. Thus the
quartic coupling of the Higgs boson is a free parameter, and there
is no expectation that it must be completely generated radiatively
(as is the case in composite Higgs models). One may be worried
whether the quartic coupling induced by mixing through heavy
composite Higgs states would be too large and thus a fine-tuning
would be needed to achieve a small value. However, the induced
quartic coupling is rendered sufficiently small by the small mixing,
thus no additional fine-tuning is needed here. Additional
contributions to the Higgs boson quartic can be induced after
integrating out heavy resonances. For example, integrating out $T'$
(through a box diagram) can induce a correction to $\lambda_H$
around $N_c y_t^4/16\pi^2$, which is also subdominant to the
observed value. Thus one does not need to worry about additional
fine tuning induced through Higgs quartic term.

Heavy composite vector bosons, transforming in adjoint
representation of SM gauge group, can be formed by the partons of
the strong dynamics. These vector bosons can mix with SM gauge
bosons. After redefinition of the SM gauge boson, such mixing will
induce a small coupling between SM charged particles to composite
vector bosons. Similar to the $H$ and $\Phi_H$ mixing, the the
mixing between gauge boson and heavy composite vector fields can be
estimated as $\epsilon\sim \frac{g_i g_*N_C}{16\pi^2\sqrt{N}}$.
After a field redefinition, the coupling between particles charged
under the SM  and heavy vectors is $\frac{g_i^2
g_*N_C}{16\pi^2\sqrt{N}}$. Integrating out these heavy vectors,
introduces dimension 6 operators with Wilsonian coefficients as
$\frac{g_i^4 g_*^2N_C^2}{(16\pi^2)^2 N\Lambda_C^2}\mathcal {O}_6$.
For $SU(3)_c$, these operators are 4-quark interactions. Given
$\Lambda_C$ higher than TeV, these operators are far from being
probed. On the other hand, the mixing between $SU(2)_L$ gauge boson
and heavy vector can be important since that can induce additional
contributions to operators like $Z\bar b b$ coupling or
$T$-parameter, for example. However, these additional contributions
are  small compared to the ones we have discussed in the previous
section as long as $\frac{g_i^4 g_*^2 N_C^2}{16\pi^2 N}<1$.
%^, which can be easily
%satisfied.

\subsection{Summary on phenomenological constraints}\label{PhenoSummary}

Before closing this section, let us summarize the most stringent
constraints in our scenario (all at 95\% confidence):

$\bullet$ The $S$-parameter imposes the strongest direct constraint on
$\Lambda_C \gsim
\sqrt{\eta_S^\Lambda}\ 1200\
\textrm{GeV}$.

$\bullet$ $Z\to b\bar b$ constrains $f\gsim \sqrt{\eta_b^{loop}}\
410 $ GeV.

$\bullet$ The LHC 4-top searches directly constrain $f\gsim
\sqrt{\eta_{4f}}\ 450 $ GeV.

$\bullet$ Assuming all mixing is from up-type sector, CP violation
in D-meson mixing gives the strongest constraint on
$f \gsim \sqrt{\eta_{4f}}\ 810 $ GeV. A $40\%$
cancellation among operators can make it
comparable to those from direct 4-top LHC searches.

The first constraint directly applies to $\Lambda_C$, all other
constraints are imposed on the suppression scale of 4-fermion
operators induced by strong dynamics. In NDA, $\lambdac \simeq 4 \pi
f$, but there are two subtleties when translating the constraints on
$f$ to $\Lambda_C$. First, it depends on the choice of strong
coupling $g_{\ast}$. Second, the representations under
$SU(N)$ of the top's partons may introduce non-trivial
$N$-dependence to 4-fermion operators. With these subtleties in
mind, we conclude the current constraints on composite scale are
$\Lambda_C\gsim $ TeV -- though if the NDA estimate were to hold
this would be closer to 4 TeV (8 TeV from mixing in the charm
sector). If $\Lambda_\psi$ is not too large and  $\Lambda_C \approx$
TeV, the fine tuning in our scenario is a few percent.

\section{Comments on Collider Signatures}
\label{sec:collider}

At a hadron collider such as the LHC, the colored parton(s)
comprising $\phi_c$ can be directly produced. If the collision
energy of a particular event is below $\Lambda_{C}$, top and bottom
quarks are the only light degrees of freedom energetically
accessible. Colored parton production is effectively the production
of the third generation quarks. One can in principle study
differential distributions, such as $d\sigma/d m_{\bar t t}$ and
$d\sigma/d m_{\bar b b}$. The third generation quarks' couplings
to gluons will acquire a from factor, and deviations in the top
quark production might be expected as the center of mass energy
approaches $\lambdac$. However, uncertainty in modeling of SM top
production is a challenge. More promising are
searches for 4-top production as outlined in the previous section.

On the other hand, if the collision energy is above $\Lambda_{C}$,
colored partons in composite top can be produced as elementary
particle, and we start to directly probe new physics of $SU(N)$
group.

First of all, we expect several heavy composite states whose masses
are about $\Lambda_{C}$, such as a $\rho'$. These particles may or
may not be charged under $SU(3)_C$, depending on their parton
content. The most straightforward way to look for such particles may
be through resonance searches. For example, one can look for the
$\rho'$ through ditop resonance search. Such analysis has been
carried out in CMS \cite{Khachatryan:2015sma}. They have searched
for a Kaluza-Klein gluon with width $\approx 15\%$ of its mass, and
find a limit of $\approx 2.3$ TeV.   We would expect there might be
color octet resonances in this scenario as well. However, as we will
see, the both the production rate and width can differ from the
target of the CMS search, thus induce a weaker constraint.

%More explicitly,
%the couplings of KK-gluon are assumed to be about 0.2$g_S$ for most
%quarks and $g_S$ for $t_R$.}

In order to compare the production cross section, we consider two
possible production channels. First, as discussed in Sec.
\ref{PhenoMixing}, a coupling between light quarks and heavy vector
resonance can be introduced through mixing with SM gluon. After
field redefinition, the coupling is about $\frac{g_S^2 g_*
N_C}{16\pi^2\sqrt{N}}$.  For comparison, in the CMS search, the
couplings of KK-gluon are assumed to be about 0.2$g_S$ for most
quarks and $g_S$ for $t_R$. If $g_*$ is mildly smaller than $4\pi$,
with a generic choice of $N$, the production cross section of
$\rho'$ can be comparable to KK-gluon considered in CMS search.

Production via gluon fusion is also possible in principle. If
the $\rho'$ is somewhat smaller than $\Lambda_C$, a rough estimate
of gluon fusion production can be obtained by integrating out the
colored states of the strong dynamics that are present above
$\Lambda_C$. Alternately, if the mass of the $\rho'$ is somewhat
larger than $\Lambda_C$, one can consider the production of the
colored partons, and estimate the production of the $\rho'$ by
considering the overlap of the partons in the $\rho'$ bound state.
Either estimate yields a production rate subdominant to the
production via the quarks discussed above, so we can expect $\rho'$
production to be comparable or smaller to that of the CMS KK gluon
benchmark.
%\reply{Furthermore, in our scenario, since all third generation
%quarks are composite, we expect democratic decay to top and bottom,
%which slightly weakens the limit from CMS.}
Moreover, for this search
to be effective, the width of the heavy resonance cannot be too
large, or else the features of a resonance are smeared in SM
background (though with precise modeling of $\bar{t}{t}$ production,
it might be possible to observe a broad excess in the future).
This likely occurs here due to the strong  $SU(N)$
interactions. Thus, direct resonance search constraints may be
comparable to, and quite possibly weaker than the indirect bounds as
found in Sec. \ref{sec:constraints}. If the resonance happens to be
somewhat narrow, resonance searches of this type could indeed be a
useful way to test this scenario, but the narrowness is not
guaranteed, as we now discuss.

One may ask whether a large width for these heavy resonances is in
contradiction with the desire to have $\Lambda_C$ and $f$ separated
by a factor smaller than $4\pi$. (Recall, the strongest bounds are
on $f$, so reducing this factor will reduce the fine-tuning in our
model, which depends directly on $\lambdac$.) Alternately, one may
wonder whether requiring $1/N$ suppression in operators suppressed
by $1/f^{2}$ (see discussion in Sec.~\ref{sec:4fermi}) will always
lead to a small decay width. We present a simple estimate to show
that a modest suppression of these operators beyond $1/f^{2}$ is
consistent with a large decay width (e.g. if $N$ is not too large).
For simplicity, we assume the only light composite particles are the
top and bottom. If there were additional (possibly SM neutral) light
composites, this could further increase the decay width heavy
resonances.

First, assume the $SU(3)_c$ colored parton is in the
fundamental representation of $SU(N)$, while the other partons in
the composite top are in different representations. Similar to the
argument for pion interactions, the coupling between a heavy
resonance and colored parton scales as $g_*/\sqrt{N}$. Consider the
case where the heavy resonance is a vector meson $\rho'$ which is an
octet of $SU(3)_C$ (analogous to  the KK gluon).  Since there is
only one way to contract the two colored partons of this composite state with the partons in the third generation fermions, its coupling is
\begin{eqnarray}
 \label{VectorCoupling}
\mathscr{L}\supset i\frac{g_*}{\sqrt{N}}\rho'_\mu
\bar\Psi\gamma^\mu\Psi.
\end{eqnarray}
Integrating out the $\rho'$ generates the 4-fermion operator at low energy,
\begin{eqnarray}
 \label{Zp4f}
\mathscr{L}\supset
\frac{g_*^2}{2 N}\frac{1}{\Lambda_C^2}(\bar\Psi\gamma^\mu\Psi)(\bar\Psi\gamma_\mu\Psi).
\end{eqnarray}
Here we have set $m_{\rho'} \simeq \lambdac$. One can further
calculate the $\rho'$ width induced by this coupling at tree level.
Including both top and bottom decay channels, we get
\begin{eqnarray}
 \label{rhoWidth}
\Gamma=\frac{g_{\ast}^2}{12 \pi N}M_{\rho'}.
\end{eqnarray}
To evade the constraints from ditop resonance searches, we require
the width of $\rho'$ be the same order as its mass
$\Gamma_{\rho'}=c\ M_{\rho'}$ where $c\sim 1$. The 4-fermion
operator induced by integrating out $\rho'$, Eq. (\ref{Zp4f}), can
be rewritten as
\begin{eqnarray}
 \label{Zp4f2}
\mathscr{L}\supset \frac{6 \pi c}{\Lambda_C^2}(\bar\Psi\gamma^\mu\Psi)(\bar\Psi\gamma_\mu\Psi).
\end{eqnarray}
At least in this simple example, we see it is consistent to
simultaneously have a wide $\rho'$ and a suppression scale modestly
larger than $\frac{\Lambda_C}{4\pi}$ in the 4-fermion interaction it
induces. If the heavy resonances are very wide, they are similar to
the $\sigma$ particle in QCD, and it could be highly non-trivial to
find these particles at a hadron collider. \footnote{One may
also be worried that $\rho'$ may be lighter than $\Lambda_C$ if its
coupling is smaller than $4\pi$ when $N$ is sizable. However, we
note that $M_{\rho'}$ does not scale with $N$. This can be easily
understood by the fact that both kinetic and mass terms of $\rho'$
are characterized by two-point correlation functions, and
$N$-dependence in the mass term is therfore removed after canonical
normalization. Moreover, if $M_{\rho'}$ is somewhat lighter
than $\Lambda_C$, this would not appreciably affect the ratio between the
width and mass, so we expect a search for the $\rho'$ would remain challenging. }

We also expect excited states of the new composite top quark,
i.e. $T'$. Again, these heavy states are expected near $\Lambda_C$,
and their widths can be large. We expect the,$\rho'$, which couples
strongly to $T'$, to have a comparable mass. If $\rho'$ is lighter
than the excited top state, $T'$ should dominantly decay to
$\rho'+t$ and consequently lead to $3t$ or $t+2b$ in its final
states. Even if $\rho'$ is a little heavier than $T'$, due to its
large coupling to $T'$, the 3-body decay channel through off-shell
$\rho'$ can still be comparable or even larger than electroweak
decay channels, such as $t+Z$, $b+W$ and  $t+H$ where $W$ and $Z$
are mainly in their longitudinal modes. The other possible decay
channel is through $t+g$. The current constraints from those
channels are around 800 GeV \cite{Chatrchyan:2013oba,Khachatryan:2015axa,ATLAS-CONF-2016-013,ATLAS:2016qlg}, thus weaker than those from EWPTs and
flavor/CP measurements discussed above. Furthermore, the low energy spectrum from $SU(N)$ strong dynamics
may not be limited to top and bottom. If there are other light
composite states, which may well be SM neutral, their existence can
change collider signatures dramatically.

This is very different from the fermionic top
partner $T$, in composite Higgs models. In those models, $T$'s have
masses much lower than confinement scale, thus there are no
additional states from strong dynamics to which they can decay.  Thus, it is more certain  top partners in these models will decay as
 $t+Z$, $b+W$ and $t+H$ (though it is possible that there might be additional light scalars in the spectrum \cite{Kearney:2013oia}).
It is also interesting to note that the $T'$ of our scenario  could conceivably
be observed but would not be responsible for the cancellation of the
ultimate quadratic divergence.  Rather they could be partially
responsible for the form factor that cuts off the first divergence
shown in Eq.~(\ref{eq:QuadHiggs}).

In summary, there are multiple reasons that new particles might be difficult to see at colliders.  First, some resonances  may be difficult to find due to their likely width, e.g. the $\rho'$ and, potentially, excitations of the top.   The states ultimately responsible for the cancellation of the final quadratic divergence near $\Lambda_{\psi}$, on the other hand, would not necessarily be expected to be wide.  For example, in the case of a SUSY UV completion, a stop could decay to a Bino and top only via a weak coupling.   Nevertheless, searches for these particle are also difficult due to the screening effect.

At sufficiently high energies additional interesting phenomena may
appear. If a pair of $SU(3)_c$ colored partons are produced, each
with energy much higher than $\Lambda_{C}$, they will produce
$SU(N)$ singlets via hadronization. It is possible that multiple
third generation quarks will be produced in the final states. The
hadron multiplicity has been studied in detail in the context of SM
QCD. For $e^+ e^-$ annihilation at center of mass energy $\sqrt s$,
the average multiplicity of any hadron species from QCD
hadronization can be approximated as \cite{Webber:1994zd}
\begin{eqnarray}
 \label{HadronNum}
\langle n(s)\rangle\sim exp\bigg\{\frac{1}{b}\sqrt{\frac{2
N_c}{\pi\alpha_s(s)}}\bigg\},
\end{eqnarray}
where $b$ is related to the beta function of strong coupling as
$\beta(\alpha_s)=-b \alpha_s^2+...$. We expect a similar expression
to control the hadronization of the $SU(N)$ group. The value of $b$
in for the new gauge group depends on the details of UV completion.
If the running of $SU(N)$ is not too fast, i.e. $b$ is not too
large, there can be an enhancement of top and bottom quark
multiplicities when the collision energy grows beyond few times
$\Lambda_{C}$.

As mentioned earlier, there could also be additional light
composite particles beyond the top and bottom.
While, if uncolored, direct production of such states is small, they could be produced in
hadronization processes, which might change the collider signatures
dramatically-- for example, giving rise to events with large missing
energy if they are stable.

\section{Conclusion}
\label{sec:conclusion} We have explored the possibility that the top
quark is composite at the TeV scale, with the Higgs boson coupling
to uncolored constituents.  Production of colored top partners at
hadron colliders can be screened via hardonization effects of the
new confining gauge group.

This scenario is likely tuned at the percent level, with
particularly strong bounds from EWPT and from flavor/CP
physics. The exact strength of these bounds is somewhat model
dependent. This scenario should be well tested at the next run of
the LHC, where effects should show up in the four top final state.
As mentioned, flavor physics represents a strong challenge for this
approach, and without careful model building, deviations would have
been expected to show up already in the kaon or $B$-meson mixing. When quark mixing arises from the up-type quarks, constraints are more modest,
and new physics is expected to arise as CP violation in the $D$-meson system, though cancellations between operators could postpone the appearance of a signal.

This model differs from the related approach of CFT duals of
Randall-Sundrum (RS) model building. For example, we have
entertained the possibility that new narrow resonances of the strong
dynamics are not present, whereas in RS scenarios, these resonances
are guaranteed as Kaluza-Klein modes on the AdS side.  Their masses and couplings are closely related to
the existence of the conformal symmetry and the assumption of large $N$ needed for the weakly coupled gravity dual.  Without these assumptions there is no assurance that these states are narrow. In addition, in the RS scenario, flavor originates from a
marriage of overlaps of extra-dimensional wave functions and Yukawa
couplings.  In our scenario, there is no extra-dimensional picture
that allow calculability of the top/bottom Yukawa couplings.

There are important differences with respect to traditional
composite Higgs (CH) theories as well. In traditional CH models,
there is a single scale of strong dynamics.  The Higgs boson is
light with respect to the scale of strong dynamics due to its
pseudo-Goldstone nature.  Here, the lightness of the Higgs boson is
ultimately ensured by additional physics above our initial (top)
compositeness scale. In traditional CH theories, a first signal is
often found in colored fermionic partners of the top quark, who
ensure that the UV cutoff is $f$, rather than $\lambdac$.  In our
theory, these top partners are not present below the confinement
scale, which allows their production to be screened.   Deviations
from precision electroweak observables, precisely because the Higgs
boson is elementary, go like $1/\lambdac$ rather than $1/f$, which
allows us to have a lower value of $\lambdac$.  Thus, we do not pay
a large fine-tuning price for the absence of top partners in the
low-energy effective theory below $\lambdac$.

The scenario as we have outlined it is truly a ``minimal" composite theory in terms of LHC phenomenology.  In  traditional CH models, it is natural to expect relatively narrow fermionic top partners because their mass is below the confinement scale.  In contrast, in our scenario, all heavy particles can be around or above $\lambdac$, and their widths can be naturally large due to strong coupling and therefore difficult to detect.  At present, this theory is not especially less fine-tuned than other CH models -- indeed, it is already  tuned at the few percent level --  but it would explain the absence of resonances into the future.   Instead, the likely proof of this theory would come in evidence for anomalous four top production at the LHC, perhaps soon. And if $\lambdac$
is close to the current experimental bound, collisions in excess of
this energy might reveal spectacular signatures, though these depend
on the details of the new strong dynamics.

\acknowledgments{We thank N.~Craig, S.~Dimopoulos,
G.~Marques-Tavares, and Y.~Tsai for discussions, and James Wells and
especially Kaustubh Agashe for readings of the manuscript. We also
wish to acknowledge several helpful questions from an anonymous
referee which have improved the paper. This work is supported by the
U.S. Department of Energy, Office of Science, under grant
DE-SC0007859.}

\begin{appendix}
\section{Examples of ``UV" models and Anomaly cancellation}
\label{sec:anomaly} In this appendix, we write a ``partial UV" model
which can induce composite third generation quarks. Anomaly
cancellation is generically a concern because the $SU(3)_c$ and EW
charge assignments in our proposal differ from those in SM. In this
section, we present two models where cancellations of all gauge
anomalies can be achieved and relevant global anomalies are matched.
We say that we present a ``partial UV" model, because our models
include a scalar $\phi_{c}$, which we do not envision as
fundamental. Our (admittedly strong) assumption is that whatever
additional dynamics gives rise to this scalar does not introduce
additional anomalies. This is also the underlying assumption applied
in Ref.~\cite{Abbott:1981yg}.

\subsection{The simplest module}
The matter content with the simplest setup is  written in Table
\ref{tab:Content}.  We take the strong group as $SU(N)$. The $SU(N)$
gauge singlets after confinement are identified with SM third
generation quarks. More explicitly, $Q_{L,3}^\alpha \sim
(\phi_c\psi_L^\alpha)$ and $t_{R}^\alpha/b_{R}^\alpha \sim
(\phi_c^\dag\psi_{R,t/b}^\alpha)$. In the IR, the theory is assumed
to be the SM, so all gauge anomalies are canceled there. In Table.
\ref{tab:Anomaly}, we show anomalies in UV theory. All anomalies can
be canceled by taking $Nx=1$.

\begin{table}[t]
\begin{center}
\begin{tabular}{ |c|c|c|c|c|}
\hline\hline \rule{0pt}{1.2em} \textrm{Particles }  & $SU(N)$
&$SU(3)_c $    & $SU(2)_L$ & $U(1)_Y$     \\
\hline $\phi_c$&$\bar{\square}$&$\square$&$\bullet$& $(\frac{1}{6}-\frac{x}{2})$ \\
\hline $\psi_L^\alpha$&${\square}$&$\bullet$&$\square$&$\frac{x}{2}$ \\
\hline $\psi_{R,t}^{\alpha}$&$\bar \square$&$\bullet$&$\bullet$& $-\frac{1}{2}(x+1)$\\
\hline $\psi_{R,b}^{\alpha}$&$\bar \square$&$\bullet$&$\bullet$&$-\frac{1}{2}(x-1)$\\
\hline\hline
\end{tabular}
\caption{The gauge charges of partons in composite third generation
quarks.} \label{tab:Content}
\end{center}
\end{table}

\begin{table}[t]
\begin{center}
\begin{tabular}{ |c|c||c|c|}
\hline\hline
$U(1)_Y$ &0&$U(1)_Y^3$&$\frac{3}{4}(1-Nx)$\\
\hline $U(1)_Y\times SU(2)_L^2$ &$\frac{1}{2}(Nx-1)$&$U(1)_Y\times SU(N)^2 $&0\\
\hline $SU(3)_c^3$ &0&$SU(N)^3$&0\\
\hline\hline
\end{tabular}
\caption{The gauge anomalies can be canceled properly in this
``partial UV" model by choosing $Nx=1$.} \label{tab:Anomaly}
\end{center}
\end{table}

Now, consider t'Hooft global anomaly matching \cite{'tHooft:1979bh}.
When $SU(N)$ becomes strongly coupled, other gauge couplings may be
treated as perturbations, and there is an approximate global
symmetry: $SU(2)_L\times SU(2)_R\times U(1)_V$. $SU(2)_L$ is
identified with the $SU(2)_L$ gauge symmetry of the SM. The
$SU(2)_R$ rotates $\psi_{R,t}^{\alpha}$ into $\psi_{R,b}^{\alpha}$.
$U(1)_V$ is related to baryon number.
For each  $SU(2)$, there are $N$ multiplets since $\psi$'s are
(anti-)fundamental under $SU(N)$. After confinement, the fermions
surviving are the bilinear products of $\phi_c$ and $\psi^\alpha$.
The left-handed quark doublet and two right-handed quarks that form a
doublet of $SU(2)_R$ transform non-trivially under the
$SU(2)_L\times SU(2)_R$ global symmetry. However, the multiplicity
of these doublets in the IR is three (due
 to the three colors under $SU(3)_c$). Since there were $N$ doublets in the UV, this is naively problematic from anomaly matching point
of view. However, particular to $SU(2)$, there is no $SU(2)_{L/R}^3$
anomaly. The only non-trivial global symmetries to be matched are
$U(1)_V\times SU(2)_L^2$ and $U(1)_V\times SU(2)_R^2$ anomalies. One
can explicitly calculate these anomalies in both UV and IR theories,
\begin{eqnarray}
 \label{GlobalAnomaly}
A_{U(1)_V\times SU(2)_L^2}|_{UV} &=& N Q_{V,\psi_L^\alpha} \nonumber\\
A_{U(1)_V\times SU(2)_L^2}|_{IR} &=& 3
(Q_{V,\psi_L^\alpha}+Q_{V,\phi_c})
\end{eqnarray}
Similar relations appear for the $U(1)_V\times SU(2)_R^2$ anomaly as
well. To match the global anomalies, we choose
$Q_{V,\phi_c} = (\frac{N}{3}-1) Q_{V,\psi_L^\alpha}=-(\frac{N}{3}-1)
Q_{V,\psi_{R,t/b}^\alpha}$. Identifying $U(1)_V$ with $U(1)_B$, we
have $Q_{V,\psi_L^\alpha}=-Q_{V,\psi_{R,t/b}^\alpha}=\frac{1}{3}$,
which indicates $Q_{V,\phi_c} = \frac{(N/3-1)}{3}$.  Note these assignments allow cancellation of the $U(1)_V^3$ anomaly

The other possibility for a consistent anomaly matching of global
symmetries is to assume the $SU(N)$ confinement spontaneously breaks
$U(1)_V$ while keeping $SU(2)_L\times SU(2)_R$ unbroken. In this
case, the anomaly matching for global symmetries is trivial since
only $SU(2)$ groups are involved. One might worry
that such a scenario may suffer from constraints on nucleon decay
since $U(1)_V$ is related to baryon symmetry and is spontaneously
broken. However, the nucleon decay rate
depends sensitively on the baryon number carried by the condensate
which spontaneously breaks $U(1)_V$. As illustrated in
Ref.~\cite{Carone:1995pu}, the nucleon decay rate is dramatically
suppressed if $U(1)_B$ is broken by a condensate which carries a
large baryon number. Depending on whether the baryon symmetry is
weakly gauged, there will be a light $U(1)_B$ gauge boson or a
Nambu-Goldstone boson in low energy spectrum. The phenomenology is model dependent and we will not discuss it any
further.

\subsection{Another approach}
As shown above, the simplest version of the ``partial UV" completion
requires either $\phi_c$ carrying baryon number or baryon number
being spontaneously broken during the confinement. It is possible
that this requirement might place non-trivial constraints on the
final UV completion.  Some additional model building can
evade these requirements.

In this case, the particle content and their charge assignments are
given in Table. \ref{tab:Content2}.
\begin{table}[t]
\begin{center}
\begin{tabular}{ |c|c|c|c|c|c|}
\hline\hline \rule{0pt}{1.2em} \textrm{Particles }  & $SU(N)$&
$SU(N+3)_1$&$SU(N+3)_2 $    & $SU(2)_L$ & $U(1)_Y$     \\
\hline $\phi_c$&$\bullet$&$\square$&$\bar{\square}$&$\bullet$& 0 \\
\hline $\psi_L^\alpha$&$\bullet$&${\square}$&$\bullet$&$\square$&$\frac{1}{6}$ \\
\hline $\psi_{R,t}^{\alpha}$&$\bullet$&$\bar \square$&$\bullet$&$\bullet$& $-\frac{2}{3}$\\
\hline $\psi_{R,b}^{\alpha}$&$\bullet$&$\bar \square$&$\bullet$&$\bullet$&$\frac{1}{3}$\\
\hline $\psi_L^{'\alpha}$&$\bar{\square}$&$\bullet$&$\bullet$&$\square$&-$\frac{1}{6}$  \\
\hline $\psi_{R,t}^{'\alpha}$&$ \square$&$\bullet$&$\bullet$&$\bullet$& $\frac{2}{3}$\\
\hline $\psi_{R,b}^{'\alpha}$&$\square$&$\bullet$&$\bullet$&$\bullet$&$-\frac{1}{3}$\\
\hline $\Delta$&$\square$&$\bullet$&$\bar \square$&$\bullet$&0\\
\hline\hline
\end{tabular}
\caption{A more complicated setup for ``partial-UV" model.}
\label{tab:Content2}
\end{center}
\end{table}
Here $SU(N+3)_1$ is the gauge group which runs strong in the high
energy. After confinement, the fermions in IR are bilinear products
of $\phi_c$ and $\psi^\alpha$. These fermions transform as
fundamental representations of an $SU(N+3)_2$ gauge group. $\Delta$ is
a Higgs field which transforms as a bilinear under
$SU(N)$ and $SU(N+3)_2$. $\Delta$ condenses at an energy a little
bit lower than the confinement scale of $SU(N+3)_1$, breaking
$SU(N)\times SU(N+3)_2$ to $SU(N)_D \times SU(3)_C$. This
condensation also will pair up the composite fermions coming from $SU(N+3)_1$
confinement with $\psi'^\alpha$ partners. The $SU(N)_D$ can be further Higgsed down to the SM gauge groups in IR, without changing the
fermionic particle content. The unpaired fermions transform
precisely as SM fermions under SM gauge groups.

The cancellation of gauge anomalies in the ``UV'' is easy to check.
Since fermions are vector-like from the $SU(N)$ and $SU(N+3)_{1,2}$
point of view, gauge anomalies of these groups are canceled.
Further, comparing the fermion content to the  fermion content of
with SM, we note that  extra particles are vector-like from the SM
gauge group point of view. Thus, the gauge anomalies of SM gauge
groups are also canceled. Finally, there is approximate
$SU(2)_L\times SU(2)_R\times U(1)_V$ global symmetry, due to
$SU(N+3)_1$ being strongly coupled. The number of fermionic
multiplets charged under $SU(2)_L\times SU(2)_R\times U(1)_V$ global
symmetry remains the same after confinement, thus the global symmetry
anomalies also match, this time without charging $\phi_c$ under
$U(1)_V$ or $U(1)_Y$.

\section{Flavor constraints from mixing in the down quark sector}

In the text, we already discussed the dangers of having mixing in
the down quark sector, as illustrated by B-meson mixing.  Here, we
enumerate additional constraints.

First consider additional constraints from the B decays.  We can induce $b \to s \gamma$ decays via the operator
\begin{equation}
{\mathcal O_7} = \frac{e m_{b} \theta_{L}^{s}}{\lambdac^2} (\bar{b} \sigma^{\mu\nu} P_{L} s) F_{\mu \nu}.
\end{equation}
This operator is particularly strongly constrained because it
interferes with the loop-induced Standard Model contribution.
Depending on the sign of the operator, this interference can be
constructive or destructive.  The experimental value is somewhat is
slightly in excess of the SM prediction, so there is a slightly
weaker constraint for the constructive case.  Adapting results of a
relatively recent evaluation \cite{Altmannshofer:2012az}, we find
\begin{equation}
\lambdac <  14  \bigl(\frac{\theta_L^s}{0.04}\bigr) \rm{TeV \, \, (constructive)}, \,\, \,   \lambdac < 54  \bigl(\frac{\theta_L^s}{0.04}\bigr) \rm{TeV \, \, (destructive)}.
\end{equation}
Given uncertainties in the relevant strong phase, measurements of
direct CP violation in $B \to K^{\ast} \gamma$ do not constrain the
imaginary part of this operator much more strongly
\cite{Altmannshofer:2014rta}. A slightly weaker constraint arises
from $B\to K^*+\mu^+ +\mu^-$\cite{Altmannshofer:2014rta}. Moreover,
due to the possibilities of anomalies in the data,  one  should take
care in setting a bound from this channel.  Indeed, it might be possible to partially explain the
anomalies (though not lepton non-universality) for particular choices of quark mixings, though
we do not pursue this further.

We may also consider $\Delta S=2$ processes, i.e. kaon oscillation
\cite{Calibbi:2012at}. These processes are suppressed by additional
mixing factors when compared to the $b$ sector. For example, the
real part of the operator with all left-handed fermions,
$\frac{1}{f^2}(\bar\Psi_L\gamma^\mu\Psi_L)(\bar\Psi_L\gamma_\mu\Psi_L)$,
gives the constraint:
\begin{eqnarray}
 \label{KOsc1}
f&\gsim & 0.17 \bigl(\frac{\theta_{L}^d}{0.004}\bigr)\bigl(\frac{\theta_{L}^s}{0.04}\bigr)\ \textrm{TeV}.
\end{eqnarray}
The constraint coming from $\epsilon$ constrains the imaginary part of the relevant operator to be smaller by a factor of $\approx$ 250.
If left-handed and right-handed mixing angles are comparable to each
other, the strongest constraints are imposed on operators of mixed chirality, i.e. $\frac{1}{f^2}(\bar\Psi_L\Psi_R)(\bar\Psi_L\Psi_R)$.
We interpret these constraints as:
\begin{eqnarray}
 \label{KOsc2}
f&\gsim & 2.3 \sqrt{\bigl(\frac{\theta_{L}^d}{0.004}\bigr)\bigl(\frac{\theta_{R}^d}{0.004}\bigr)\bigl(\frac{\theta_{L}^s}{0.04}\bigr)\bigl(\frac{\theta_{R}^s}{0.04}\bigr)}\ \textrm{TeV,}
\end{eqnarray}
again with correspondingly stronger bounds when a phase
is present.  For general mixings, these constraints may compete with
the constraints from $B$ meson mixing presented in the text, though
it depends on the precise choice of phase, which in the general
mixing case is not locked to the CKM phase.

\end{appendix}

\bibliography{tc}{}

\end{document}